\documentclass[11pt]{article}

\usepackage{cite}
\usepackage[cmex10]{amsmath}
\usepackage{algorithm}
\usepackage{algorithmic}

\usepackage{array}

\usepackage{mdwmath}
\usepackage{mdwtab}

\usepackage{eqparbox}

\usepackage[font={small}]{caption}
\usepackage{fixltx2e}

\usepackage{stfloats}

\usepackage{url}

\usepackage{amsmath}
\usepackage{amssymb}
\usepackage{amsmath}
\usepackage{amsmath,amssymb,euscript,mathrsfs,amsthm}
\usepackage{graphicx,color,subfig}
\usepackage{tikz}
\usepackage{enumerate}

\usetikzlibrary{automata,arrows,shapes,snakes,automata,backgrounds,petri}
\usepackage[latin1]{inputenc}
\usepackage{verbatim}
\usepackage{color}

\newcommand{\cref}[1]{Claim~\ref{#1}}

\newtheorem{theorem}{\bf Theorem }

\newtheorem{lemma}{\bf Lemma}

\newcommand{\vect}[1]{\boldsymbol{#1}}
\newcommand{\mat}[1]{\boldsymbol{#1}}
\newcommand{\SET}{\mathbb{S}}
\newcommand{\CMP}{\mathbb{C}}
\newcommand{\REAL}{\mathbb{R}}
\newcommand{\BIGO}{\mathcal{O}}
\newcommand{\IMG}{{\bf i}}
\newcommand{\TSP}{^{\rm T}}
\newcommand{\HET}{^{\rm H}}
\newcommand{\DIAG}[1]{{\rm diag}(#1)}
\newcommand{\PRO}[1]{\mathbb{P}\left\{#1\right\}}
\newcommand{\PROL}[1]{\mathbb{P}\{#1\}}
\newcommand{\EXP}[1]{\mathbb{E}\left[#1\right]}
\newcommand{\EXPL}[1]{\mathbb{E}[#1]}
\newcommand{\SEP}[1]{\left\|#1\right\|_{{\it \psi}_1}}
\newcommand{\SEPL}[1]{\|#1\|_{{\it \psi}_1}}
\newcommand{\LNR}{\mathcal{A}}
\newcommand{\SGS}[1]{\left\|#1\right\|_{{\it \psi}_2}}
\newcommand{\SGSL}[1]{\|#1\|_{{\it \psi}_2}}
\newcommand{\ABS}[1]{\left|#1\right|}
\newcommand{\ABSL}[1]{|#1|}
\newcommand{\FBN}[1]{\left\|#1\right\|_F}
\newcommand{\FBNL}[1]{\|#1\|_F}
\newcommand{\OPN}[1]{\left\|#1\right\|}
\newcommand{\OPNL}[1]{\|#1\|}
\newcommand{\TWON}[1]{\left\|#1\right\|_2}

\newcommand{\ONEN}[1]{\left\|#1\right\|_1}

\newcommand{\TR}[1]{{\rm tr}\left(#1\right)}

\newcommand{\SUPP}[1]{{\rm supp}(#1)}

\renewcommand\footnoterule{\kern-3pt \hrule width 2in \kern 2.6pt}

\definecolor{myblue}{RGB}{80,80,160}
\definecolor{mygreen}{RGB}{80,160,80}

\title{Fast and Robust Compressive Phase Retrieval with Sparse-Graph Codes}

\author{Dong Yin, Kangwook Lee, Ramtin Pedarsani, Kannan Ramchandran\\
Department of Electrical Engineering and Computer Sciences\\
University of California, Berkeley\\
Email: $\{$dongyin, kw1jjang, ramtin, kannanr$\}$@eecs.berkeley.edu}

\begin{document}
\maketitle

\begin{abstract}
In this paper, we tackle the compressive phase retrieval problem in the presence of noise. 
The noisy compressive phase retrieval problem is to recover a $K$-sparse complex signal $\vect{s} \in \CMP^n$,
from a set of $m$ noisy quadratic measurements: $ y_i=\ABSL{\vect{a}_i\HET\vect{s}}^2+w_i,$
where $\vect{a}_i\HET\in\CMP^n$ is the $i$th row of the measurement matrix $\mat{A}\in\CMP^{m\times n}$, and $w_i$ is the additive noise to the $i$th measurement. We consider the regime where $K=\beta n^\delta$, $\delta\in(0,1)$.
We use the architecture of {\em PhaseCode} algorithm \cite{Phasecode}, and robustify it using two schemes: the almost-linear scheme and the sublinear scheme. We prove that with high probability, the almost-linear scheme recovers $\vect{s}$ with sample complexity\footnote{
Here, we define the notations $\BIGO(\cdot)$, $\Theta(\cdot)$, and $\Omega(\cdot)$. We have $f=\BIGO(g)$ if and only if there exists a constant $C_1>0$ such that $\ABS{f/g}<C_1$; $f=\Theta(g)$ if and only if there exist two constants $C_1,C_2>0$ such that $C_1<\ABS{f/g}<C_2$; and $f=\Omega(g)$ if and only if there exists a constant $C_1>0$ such that $\ABS{f/g}>C_1$.}
$\Theta(K \log(n))$ and computational complexity $\Theta(n \log(n))$, and the sublinear scheme recovers $\vect{s}$ with sample complexity $\Theta(K\log^3(n))$ and computational complexity $\Theta(K\log^3(n))$. To the best of our knowledge, this is the first scheme that achieves sublinear computational complexity for compressive phase retrieval problem. Finally, we provide simulation results that support our theoretical contributions.
\end{abstract}


%

\section{Introduction}\label{sec:intro}

\subsection{Problem Formulation}
In this paper, we consider the noisy compressive phase retrieval problem.
The noisy compressive phase retrieval problem is to recover a sparse complex signal $\vect{s}$,
from a set of quadratic measurements 
$$
y_i=\ABS{\vect{a}_i\HET\vect{s}}^2+w_i,~~i\in[m],
$$
where $\vect{a}_i\HET\in\CMP^n$ are rows of the measurement matrix $\mat{A}\in\CMP^{m\times n}$, $w_i$'s are noise, and $[m]$ denotes the set $\{1,2,\ldots,m\}$. We assume that $w_i$'s are independent, zero-mean, sub-exponential \cite{nonasym} random variables. This model is considered in many phase retrieval literatures \cite{Phaselift, Polar, SquareOutput}. As mentioned in \cite{Phaselift}, in many applications such as optics \cite{bunk2007diffractive}, one can measure squared-magnitudes rather than magnitudes. Our goal is to design $\mat{A}$ and recover $\vect{s}$ up to a global phase from the $y_i$'s with small sample and computational complexity.
Although the measurement matrix cannot be freely designed in some cases \cite{loewen1997diffraction}, considering the most general compressive phase retrieval problem can provide the insight to tackle more constrained problems. Moreover, there is no constraint on the design of $\mat{A}$ in some applications such as quantum optics \cite{mirhosseini2014compressive}.

We also assume that signal $\vect{s}$ is quantized, which means that the components of $\vect{s}$ lie in a finite set of complex numbers. 
More specifically, let $L_m$ and $L_p$ be the number of possible magnitudes and phases of the non-zero components, respectively, and each component of $\vect{s}$ is in the set 
$$\SET=\{u\varepsilon e^{\IMG \frac{2\pi (v-1)}{L_p}} | u\in[L_m], v \in[L_p] \}\cup\{0\}\subset\CMP,$$
where $\varepsilon>0$ and $\IMG$ denotes the imaginary unit. Quantized signals can be good approximations of the real world signals and are natural for signal processing with computers \cite{love2004value, candy1974use}. Additionally, we assume $\vect{s}$ is $K$-sparse\footnote{We define the \emph{support}, denoted by $\SUPP{\vect{s}}$, to be the set of the indices of the non-zero components of $\vect{s}$.}, i.e., $\ABS{\SUPP{\vect{s}}}=K$.  
In this paper, we consider the regime where there exist two constants $\beta$ and $\delta$ such that $K=\beta n^\delta$, $\delta\in(0,1)$. 

\subsection{Main Contributions}
In this paper, we propose two schemes: \emph{almost-linear} and \emph{sublinear} schemes for noisy compressive phase retrieval. These two schemes are robust versions of the PhaseCode algorithm \cite{Phasecode}, which is a fast and effective framework for the noiseless scenarios. 
The key idea of PhaseCode is the usage of sparse-graph codes, a powerful tool from coding theory. Sparse-graph codes have been widely applied in communications \cite{richardson2008modern} and signal processing \cite{li2014sub,pawar2013computing}.
The main advantage of our schemes is the small sample and computational complexity\footnote{In this paper, we set $\log(n)$ to be log base 2.}, as shown in Table \ref{sample_computation}.
\begin{table}[!h]
\caption{Sample and computational complexity}
\label{sample_computation}
\centering
\begin{tabular}{ccc}
\hline
 & almost-linear & sublinear\\
\hline
sample complexity &$\Theta(K\log(n))$ & $\Theta(K\log^3(n))$\\
computational complexity &$\Theta(n\log(n))$ & $\Theta(K\log^3(n))$\\
\hline
\end{tabular}
\end{table}

The sublinear scheme uses slightly more samples than the almost-linear scheme but the computational complexity is much smaller. To the best of our knowledge, the sublinear scheme is the first proposed algorithm that achieves sublinear computational complexity in the signal dimension $n$ for compressive phase retrieval problem.



\section{Related Work}\label{sec:related_work}

\subsection{Previous Works on Robust Phase Retrieval}
The phase retrieval problem has been studied extensively over several decades.  We do not attempt to provide a comprehensive  literature review here; instead, we highlight only some of the pertinent and diverse approaches to this problem that we are aware of.  There are two popular classes of approaches, one based on convex-optimization methods, and the other based on greedy methods such as gradient descent and alternation minimzation.
In the first class, the bulk of the literature on phase retrieval problems is dedicated to the non-sparse signal regime, where the signal has no sparsity-structure to be exploited. ``Phaselift" \cite{Phaselift} and ``PhaseCut"  \cite{Waldspurger} are seminal examples
of this class, featuring the use of convex relaxation methods based on Semi-Definite Programming (SDP).  While SDP-based algorithms can provide provable performance guarantees and are robust to noise, they typically suffer from prohibitively high computational and memory complexity.   
There are also interesting works on the use of SDP-based
approaches to exploit signal sparsity in the compressive phase retrieval \cite{SquareOutput, Li,Hassibi1,Hassibi2}. 
The second class of methods, which are popular in practice, is based on greedy methods.  In general, these algorithms have a reasonable computational complexity, and are therefore used in many practical applications \cite{gerchberg}.  However, with the exception of a few recent works  \cite{Sanghavi,candes2014phase}, this class of algorithms generally comes with little theoretical guarantees.  

\subsection{PhaseCode algorithm}
As mentioned in Section \ref{sec:intro}, our proposed schemes are based on the PhaseCode algorithm.
Here we briefly review the basic ideas of PhaseCode\footnote{Here, we only consider the Unicolor PhaseCode algorithm.}.

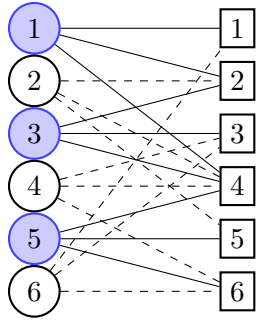
\begin{figure}[!t]
        \vspace{-0.1in}
	\centering
	\begin{tikzpicture}[node distance=0.7cm,>=stealth',bend angle=45,auto]
  	\tikzstyle{ball}=[circle,thick,draw=blue!75,fill=blue!20,minimum size=3mm]
  	\tikzstyle{emptyball}=[circle,thick,draw=black,fill=white,minimum size=3mm]
    	\tikzstyle{bin}=[rectangle,thick,draw=black,fill=white,minimum size=3mm]
  	\begin{scope}
  
      	\node [ball] (L1) {1};
    	\node [emptyball] (L2) [below of=L1] {2};
   	\node [ball] (L3)  [below of=L2] {3};
    	\node [emptyball] (L4) [below of=L3] {4};
    	\node [ball] (L5) [below of=L4] {5};
	\node [emptyball] (L6) [below of=L5] {6};
	
	\node [bin] (R1) [right of=L1, xshift=2.0cm] {1};
	\node [bin] (R2) [right of=L2, xshift=2.0cm] {2};
	\node [bin] (R3) [right of=L3, xshift=2.0cm] {3};
	\node [bin] (R4) [right of=L4, xshift=2.0cm] {4};
	\node [bin] (R5) [right of=L5, xshift=2.0cm] {5};
	\node [bin] (R6) [right of=L6, xshift=2.0cm] {6};
	
	\path (L1) edge [left] (R1);
	\path (L1) edge [left] (R2);
	\path (L1) edge [left] (R4);
	
	\path (L2) edge [left, dashed] (R2);
	\path (L2) edge [left, dashed] (R4);
	\path (L2) edge [left, dashed] (R5);
	
	\path (L3) edge [left] (R2);
	\path (L3) edge [left] (R3);
	\path (L3) edge [left] (R4);
	
	\path (L4) edge [left, dashed] (R3);
	\path (L4) edge [left, dashed] (R4);
	\path (L4) edge [left, dashed] (R6);
	
	\path (L5) edge [left] (R4);
	\path (L5) edge [left] (R5);
	\path (L5) edge [left] (R6);
	
	\path (L6) edge [left, dashed] (R1);
	\path (L6) edge [left, dashed] (R3);
	\path (L6) edge [left, dashed] (R6);
	
  	\end{scope}
	\end{tikzpicture}
	\caption{An example of PhaseCode. Bins 1, 3, 5, 6 are singletons with singleton balls 1, 3, 5, 5, which can be found in the first iteration of PhaseCode algorithm. Then, the algorithm finds a strong doubleton: bin 2, and the relative phases between balls 1 and 3. In the next iteration, the algorithm finds a resolvable multiton bin 4 and colors ball 5. After that, no more balls can be colored. The algorithm stops and successfully finds all the non-zero components.}
	\vspace{-0.15in}
	\label{fig:example}
\end{figure}

The PhaseCode algorithm iteratively recovers the non-zero components via a \emph{ball coloring} algorithm based on a ``divide-and-conquer'' philosophy. The measurement matrix of PhaseCode algorithm $\mat{A}\in\CMP^{4M\times n}$ is designed to be a row tensor product of a trigonometric modulation matrix $\mat{A}_0\in\CMP^{4\times n}$ 
and a code matrix $\mat{H}\in \{0,1\}^{M\times n}$, i.e., $\mat{A}=\mat{A}_0\otimes\mat{H}$. This means we have
$
\mat{A}=[ \mat{A}_1\HET\ \mat{A}_2\HET\ \cdots\ \mat{A}_M\HET ]\HET,
$
where $\mat{A}_i=\mat{A}_0\DIAG{\vect{h}_i}\in\CMP^{4\times n}$ and $\vect{h}_i$ is the $i$th row of $\mat{H}$. Here, $\DIAG{\vect{h}_i}$ denotes a diagonal matrix whose diagonal entries are the entries of $\vect{h}_i$. Each of the $\mat{A}_i$'s gives us a set of 4 measurements. PhaseCode's measurement system can be equivalently represented using a balls-and-bins model, or a bipartite graph model. In this representation, there are $n$ balls and $M$ bins, and the balls and bins correspond to the components of $\vect{s}$ and the sets of 4 measurements, respectively. Then, $\mat{A}_i$ is the measurement matrix of the $i$th bin, and $\mat{H}$ is the biadjacency matrix of the bipartite graph. To be specific, if $h_{ij}=1$, the $j$th ball is put into the $i$th bin. For example, in Figure \ref{fig:example}, the third bin has
$
\vect{h}_3=[0~~0~~1~~1~~0~~1].
$
If a ball corresponds to one of the $K$ non-zero components of $\vect{s}$, it is called an \emph{active ball}. And we simply choose the bipartite graph to be $d$-left regular, i.e., each ball is connected to $d$ bins chosen from the $M$ bins uniformly at random. 

We can classify bins according to the number of active balls in them. 
A \emph{zeroton} is a bin with no active balls; a \emph{singleton} is a bin with one active ball, which is called a \emph{singleton ball}; a \emph{doubleton} is a bin with two active balls; a \emph{multiton} is a bin with more than one active balls.\footnote{This implies that a doubleton is also a multiton.} We also define \emph{strong doubletons}, which are doubletons consisting of two singleton balls. For a multiton, if we know the indices, magnitudes and relative phases of all of the active balls except one, we call it a \emph{resolvable multiton}.

In the noiseless case, the 4 measurements in each bin are carefully designed so that the decoding algorithm can detect singletons, resolve strong doubletons and resolvable multitons. To be specific, the decoding algorithm can detect whether a bin is a singleton, and if it is, the decoder can find the location index and magnitude of the active ball;
if magnitudes of the two active balls in a doubleton is known, the decoder can find their relative phase; for a resolvable multiton, the decoder can calculate the index and magnitude of the unknown ball, and the relative phase between this ball and others.

The decoding algorithm of PhaseCode is an iterative process. In the first iteration, it resolves all the singletons. In the second iteration, it resolves all the strong doubletons consisting of the singleton balls found in the previous iteration and gets the relative phases between the two singleton balls in them. Then, the algorithm finds the largest set of singleton balls whose relative phases are known and call these balls \emph{colored}. In the following iterations, the algorithm iteratively checks whether the bins are resolvable multitons and colors the remaining balls. In \cite{Phasecode}, it is shown that in order to guarantee successful recovery with high probability, we need to use $M=\Theta(K)$ bins. It is proved that PhaseCode algorithm can recover a fraction $1-p$, for arbitrarily small $p$, of the non-zero elements with probability $1-\BIGO(1/K)$, with $m=\Theta(K)$ measurements\footnote{To be more specific, the authors characterized the exact number of measurements and the corresponding fraction of recoverable balls: $14K$ measurements with $p=10^{-7}$.}
and the computational complexity of the algorithm is $\Theta(K)$. The PhaseCode algorithm is illustrated by a simple example in Figure \ref{fig:example}. 

In practice, the measurements are corrupted by noise, and in this case, we can not use only 4 measurements in each bin for the decoding algorithm. 
However, we can robustify the algorithm by redesigning the measurement pattern $\mat{A}_0$ while keeping the code matrix $\mat{H}$ and the ball coloring algorithm the same as the noiseless case.

\section{Main Results}\label{sec:main}
We propose two schemes to robustify PhaseCode in the presence of noise: almost-linear scheme and sublinear scheme. The main results of this paper are the following theorems.

\begin{theorem}\label{thm_exh_overall}
The almost-linear scheme can recover a fraction $1-p$, for arbitrarily small $p$, of the non-zero elements of $\vect{s}$ with probability $1-\BIGO(1/K)$, 
with $\Theta(K\log(n))$ measurements. The computational complexity of the algorithm is $\Theta(n\log(n))$.
\end{theorem}

\begin{theorem}\label{thm_fast_overall}
The sublinear scheme can recover a fraction $1-p$, for arbitrarily small $p$, of the non-zero elements of $\vect{s}$ with probability $1-\BIGO(1/K)$,
with $\Theta(K\log^3(n))$ measurements. The computational complexity of the algorithm is $\Theta(K\log^3(n))$.
\end{theorem}
See the proofs of Theorems \ref{thm_exh_overall} and \ref{thm_fast_overall} in Appendix \ref{prf_exh_overall} and \ref{prf_thm_fast}. Details of the measurement design and the decoding algorithm are shown in the following sections.

\section{Almost-linear Scheme}\label{sec:exh_search}

The idea of the almost-linear scheme is to encode the columns as different patterns. With the number of measurements in each bin being $\Theta(\log(n))$, the patterns are guaranteed to be different enough, so that we can successfully resolve singletons or uncolored balls in resolvable multitons.
\subsection{Design of Measurements}

Instead of using the 4-by-$n$ trigonometric modulation matrix, we use a new random matrix $\mat{A}_0=\{a_{ij}\}_{P\times n}$ whose entries are i.i.d. with the following distribution:
\begin{equation}\label{distribution}
 a_{ij}=
\begin{cases}
0, & \text{with probability } 1/2 \\
e^{\IMG \theta_{ij}}, & \text{with probability } 1/2,
\end{cases}
\end{equation}
where $\theta_{ij}$'s are i.i.d. and uniformly distributed in $[0,2\pi)$. We call $\mat{A}_0$ the \emph{test matrix}, and we can show that we need $P=\Theta(\log(n))$ for each bin to achieve successful recovery. 

For the almost-linear algorithm, the measurement matrix of the $l$th bin is
$
\mat{A}_l=\mat{A}_0\DIAG{\vect{h}_l}.
$
Without loss of generality, we omit bin index $l$, and simply use $\vect{h}$ to denote the coding pattern of any bin.
Then the measurements of this bin would be
\begin{equation}\label{measurement}
y_i=\ABS{\vect{a}_i\HET \DIAG{\vect{h}}\vect{s}}^2+w_i,\ i\in[P],
\end{equation}
where $\vect{a}_i\HET$ is the $i$th row of $\mat{A}_0$, and the noise $w_i\in\REAL$, $i\in[n]$ 
satisfies the properties given in Section \ref{sec:intro}.
To simplify notation, we define a linear map $\LNR$ from $\CMP^{n\times n}$ to $\REAL^P$:
\begin{equation}\label{lineardef}
\LNR:\ \mat{X}\mapsto \{\vect{a}_i\HET \mat{X} \vect{a}_i\}_{i\in[P]}.
\end{equation}
Now according to (\ref{measurement}), by defining $\vect{x}=\DIAG{\vect{h}}\vect{s}$, we have $\vect{y}=\LNR(\vect{x}\vect{x}\HET)+\vect{w}$, 
where $\vect{y}=\{y_i\}_{i\in[P]}$ and $\vect{w}=\{w_i\}_{i\in[P]}$ are the measurement vector and noise vector, respectively. We call $\vect{x}$ the \emph{true signal} corresponding to this bin. 

\subsection{Decoding Algorithm}
As mentioned in Section \ref{sec:related_work}, PhaseCode algorithm requires the measurements in each bin to handle three operations, i.e., 
detecting singletons, resolving strong doubletons, and detecting resolvable multitons and coloring the uncolored ball in it. Using our new measurement system, these operations can be done reliably by a simple guess-and-check method: we guess all possible indices, magnitudes, and relative phases, and use an energy test to decide whether our guess is correct. For any of the three operations, we make hypothesis on the unknown index, magnitude, and phase of the true signal $\vect{x}$ and construct the corresponding hypothesis signal $\hat{\vect{x}}$. For example, when we do singleton detecting, if our hypothesis is that the bin is a singleton, and that the location index of the active ball is 5 with the magnitude being $3\varepsilon$, 
we construct $\hat{\vect{x}}=3\varepsilon \vect{e}_5$, where $\vect{e}_i$ denotes the $i$th vector of the canonical basis.
Similarly, we can resolve strong doubletons. For instance, suppose that we know a bin has two singleton balls, which are located at $2$ and $5$, respectively, and we also know the magnitudes of the two balls are $2\varepsilon$ and $3\varepsilon$, respectively. Then, if we can make a hypothesis that the relative phase is $\frac{\pi}{4}$, we can construct $\hat{\vect{x}}=2\varepsilon\vect{e}_2+3\varepsilon e^{\IMG\frac{\pi}{4}}\vect{e}_5$. Then, we need to check whether our hypothesis is correct. To do this, we perform an $\ell_1$ norm energy test shown in (\ref{energy_test}):
\begin{equation}\label{energy_test}
\begin{aligned}
\hat{\vect{x}}\sim\vect{x}&,\text{ if } \frac{1}{P}\ONEN{\vect{y}-\LNR(\hat{\vect{x}}\hat{\vect{x}}\HET)}<t_0, \\
\hat{\vect{x}}\nsim\vect{x}&,\text{ otherwise},  
\end{aligned}
\end{equation}
where $\hat{\vect{x}}\sim\vect{x}$ means $\hat{\vect{x}}$ and $\vect{x}$ are equal up to a global phase, and $t_0$ is the threshold. The intuitive reason why we do this test is that when $\hat{\vect{x}}\sim\vect{x}$, $\LNR(\hat{\vect{x}}\hat{\vect{x}}\HET)=\LNR(\vect{x}\vect{x}\HET)$, 
then $\vect{y}-\LNR(\hat{\vect{x}}\hat{\vect{x}}\HET)=\vect{w}$, whose energy should be small. 
Conversely, when $\hat{\vect{x}}\nsim\vect{x}$, the energy of $\vect{y}-\LNR(\hat{\vect{x}}\hat{\vect{x}}\HET)$ should be large. Here, we give a result on the error probability of the energy test.
\begin{lemma}\label{lem:energy_test}
When $P=\Theta(\log(n))$ and $\varepsilon$ is appropriately large, with proper threshold $t_0$, the error probability of the energy test shown in (\ref{energy_test}) is $\BIGO(1/n^2)$.
\end{lemma}
The proof of this lemma follows the similar idea which appears in Lemma 14 in \cite{chen2014convex}. We can also show that we need to perform $\Theta(n)$ energy tests before the algorithm stops. Then, using Lemma \ref{lem:energy_test} and some basic principles in probability theory, we can show that the failure probability of almost-linear scheme is $\BIGO(1/K)$. As for the sample and computational complexity, since we have $\Theta(\log(n))$ measurements in each bin and $\Theta(K)$ bins, the sample complexity of almost-linear scheme would be $\Theta(K\log(n))$;
and since the computational cost of each test is $\Theta(\log(n))$ and there are $\Theta(n)$ tests, the computational complexity of almost-linear scheme is $\Theta(n\log(n))$.

\section{Sublinear Scheme}\label{sec:fast_search}
Although the $\BIGO(n\log(n))$ computational complexity of almost-linear scheme is compelling, we can further improve the computational complexity. Recall that in the noiseless scenario, we get the location index of the singletons and the uncolored balls in resolvable multitons by only looking at the measurements. Based on this idea, we propose the sublinear scheme for the noisy scenario, which can achieve much lower computational cost compared to the almost-linear scheme, at the cost of slightly larger sample complexity.

\subsection{Design of Measurements}

In the sublinear scheme, the measurement matrix in each bin is designed to be a concatenation of the test matrix $\mat{A}_0$ defined in the almost-linear scheme and $R$ \emph{index matrices} $\mat{F}_1,\ldots,\mat{F}_R$. The test matrix $\mat{A}_0$ is still used to perform the energy tests and the index matrices are used to find the location indices. 

Now we show how to design the index matrices. The main idea is to encode each column as a binary code such that we can directly decode the column index from the measurements. The similar idea is also used in the Chaining Pursuit method\cite{chainpursuit}. First, we define a deterministic matrix $\mat{B}=\{b_{ij}\}\in\{0,1\}^{R\times n}$, where $ R=\lceil \log n \rceil$, and the $i$th column of $\mat{B}$ is the binary representation of the integer $i-1$. For example, when $n=4$, we have, 
$$
\mat{B} = \left[
\begin{array}{cccc}
0 & 0 & 1 & 1 \\
0 & 1 & 0 & 1 
\end{array}
\right].
$$
We use $\vect{b}_i$ and $\vect{B}_j$ to denote the $i$th row and $j$th column of $\mat{B}$, respectively. Let $\mat{F}_0\in\CMP^{Q\times n}$ be a random matrix whose elements are i.i.d. and uniformly distributed on the unit circle, and $\mat{F}=\mat{F}_0\otimes\mat{B}\in\CMP^{RQ\times n}$.
This means we have $\mat{F}=[ \mat{F}_1\HET \  \mat{F}_2\HET \ \cdots \mat{F}_R\HET  ]\HET$, where $\mat{F}_i=\mat{F}_0\DIAG{\vect{b}_i}\in\CMP^{Q\times n}$.
By concatenating with the test matrix, the measurement matrix of the $l$th bin is $\mat{A}_l=[\mat{A}_0\HET\ \mat{F}\HET]\HET\DIAG{\vect{h}_l}\in\CMP^{(P+QR)\times n}$. Here, we give a simple example of $\mat{A}_l$. Let $n=4$ and thus $R=2$. We have
\begin{equation}\label{ex:matrix_fast}
\mat{A}_l = 
\left[
\begin{array}{cccc}
\vect{A}_{0,1} & \vect{A}_{0,2} & \vect{A}_{0,3} & \vect{A}_{0,4} \\ \hline
\vect{0}  & \vect{0} & \vect{F}_{0,3} & \vect{F}_{0,4} \\
\vect{0} & \vect{F}_{0,2} & \vect{0} & \vect{F}_{0,4}
\end{array}
\right]\DIAG{\vect{h}_l},
\end{equation}
where $\vect{A}_{0,i}$'s and $\vect{F}_{0,i}$'s are the columns of $\mat{A}_0$ and $\mat{F}_0$. We can show that we need $Q=\Theta(\log^2(n))$ to reliably find the correct location index and we also need $P=\Theta(\log(n))$ to perform energy tests.

Consequently, there are $R+1$ sets of measurements. The first set $\vect{y}_0=\{y_{0,i}\}_{i\in[P]}$ is the same as the measurements in almost-linear scheme and is called the \emph{test measurements}:
$$
y_{0,i}=\ABS{\vect{a}_i\HET\vect{x}}^2+w_{0,i},\ i\in[P],
$$
where $\vect{x}=\DIAG{\vect{h}}\vect{s}$ and is still called the true signal. The other $R$ sets $\vect{y}_j=\{y_{j,i}\}_{i\in[Q]}$, $j\in[R]$ correspond to the index matrices and are called the \emph{index measurements}. Each set is composed of $Q$ measurements:
$$
y_{j,i}=\ABS{\vect{f}_{j,i}\HET \vect{x}}^2+w_{j,i},\ i\in[Q],\ j\in[R],
$$
where $\vect{f}_{j,i}\HET$ is the $i$th row of $\mat{F}_j$. We also let $\vect{w}_j$'s be the noise vectors, $j\in\{0\}\cup[R]$.



\subsection{Decoding Algorithm}

The sublinear scheme can find the location index by only looking at the measurements. For example, assume that the bin with measurement matrix in (\ref{ex:matrix_fast}) is a singleton whose non-zero component is at position 2. Then, the decoder can see that the elements of the first set of index measurements $\vect{y}_1$ have small absolute value since these measurements only contain noise. Now the decoder knows that the non-zero element should be in the first half of the signal. Then it sees that the elements in $\vect{y}_2$ have large energy. The decoder knows that if it is indeed a singleton, the only possible index of the non-zero component would be 2. Actually this procedure is a binary search on all the $n$ indices of the signal. After this indexing process, the decoder can use the same way as the almost-linear scheme to construct a signal $\hat{\vect{x}}$ as the hypothesis of the true signal of this bin, and then use the testing measurements to perform the same energy test.

Now we formally show the details of the fast index search. Assume that $\ABS{\SUPP{\vect{x}}}=T$, and there are $T_s$ uncolored balls in this bin. More specifically,
$\vect{x}=\vect{x}_c+\vect{x}_s$, $\ABS{\SUPP{\vect{x}_s}}=T_s$, $\SUPP{\vect{x}_c}\cap\SUPP{\vect{x}_s}=\emptyset$,
and we know a vector $\hat{\vect{x}}_c\sim\vect{x}_c$. Note that when $T=T_s=1$, we have $\hat{\vect{x}}_c=\vect{x}_c=0$.
Our goal is to find the index $l_s$ of the non-zero element in $\vect{x}_s$ when $T_s=1$ and $\SUPP{\vect{x}_s}=\{l_s\}$.
When $T=1$ and $T>1$, we are looking for singleton balls and uncolored balls in resolvable multitons, respectively. We subtract the measurements contributed by the signal components which are known. More specifically, let $\hat{y}_{j,i}=\ABSL{\vect{f}_{j,i}\HET \hat{\vect{x}}_c}^2$, and $\tilde{y}_{j,i}=y_{j,i}-\hat{y}_{j,i}$. We perform the following index tests for $j\in[R]$ with threshold $t_1>0$ to get $l_s$:
\begin{equation}\label{fast_index_test}
\begin{aligned}
\tilde{b}_j=0&,\text{ if } \ABS{\frac{1}{Q}\sum_{i=1}^Q\tilde{y}_{j,i}} <t_1, \\
\tilde{b}_j=1&,\text{ otherwise}. 
\end{aligned}
\end{equation}
The index tests output a binary string $\tilde{\vect{b}}=\{\tilde{b}_j\}_{j\in[R]}$. We should also notice that if $T_s>1$, we can still get an output after the index tests, 
but the energy test with the test measurements prevents us from making mistakes. 
Lemma \ref{lem:fast_one_section} tells us that with high probability $\tilde{b}_j=b_{jl_s}$.
\begin{lemma}\label{lem:fast_one_section}
When $Q=\Theta(\log^2(n))$, with proper threshold $t_1$, if $\SUPP{\vect{x}_s}=\{l_s\}$, then $\mathbb{P}\{\tilde{b}_j\neq b_{jl_s}\}=\BIGO(1/K^3)$.
\end{lemma}
Similar to the almost-linear scheme, using Lemma \ref{lem:fast_one_section}, we can prove that the failure probability of sublinear scheme is $\BIGO(1/K)$. Since the total number of measurements in each bin is $P+RQ=\Theta(\log^3(n))$, the sample complexity of sublinear scheme is $\Theta(K\log^3(n))$. In terms of the computational complexity, since there are $\Theta(K)$ bins and a constant number of iterations, the computational complexity of sublinear algorithm is $\Theta(K\log^3(n))$.
\section{Simulation Results}\label{sec:experiment}
In this section, we show the simulation results to support our theory. The simulations are conducted in Python. Since the sublinear scheme has much lower computational complexity than the almost-linear scheme, we only conduct simulations on the sublinear scheme here. We define the signal-to-noise ratio (SNR):
$$
\text{SNR}=10\log_{10}{\frac{\sum_{j=0}^R\TWON{\vect{y}_j-\vect{w}_j}^2}{\sum_{j=0}^R\TWON{\vect{w}_j}^2}},
$$
and use Gaussian noise. Since the fraction of unrecovered balls $p$ can be arbitrarily small, in the simulations, we simply define a successful recovery as the cases when all non-zero components are correctly found up to a global phase. In all the simulations, we set $P=5\log(n)$, $d=15$, $M=8K$, $L_m=3$, $L_p=6$, and $\varepsilon=1$.
\begin{figure}[h]
\centering
\vspace{-0.15in}
\includegraphics[width=0.7\textwidth]{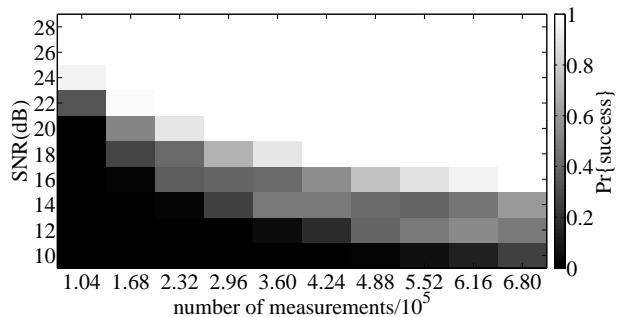}
\caption{Probability of successful recovery. We choose $n=2^{20}$ and $K=50$. Different values of $Q$ and SNR are tested, and for each set of parameters, 1000 experiments are conducted.}
\vspace{-0.2in}
\label{fig_sim_snr}
\end{figure}
\begin{figure}[h]
\centering
\includegraphics[width=0.7\textwidth]{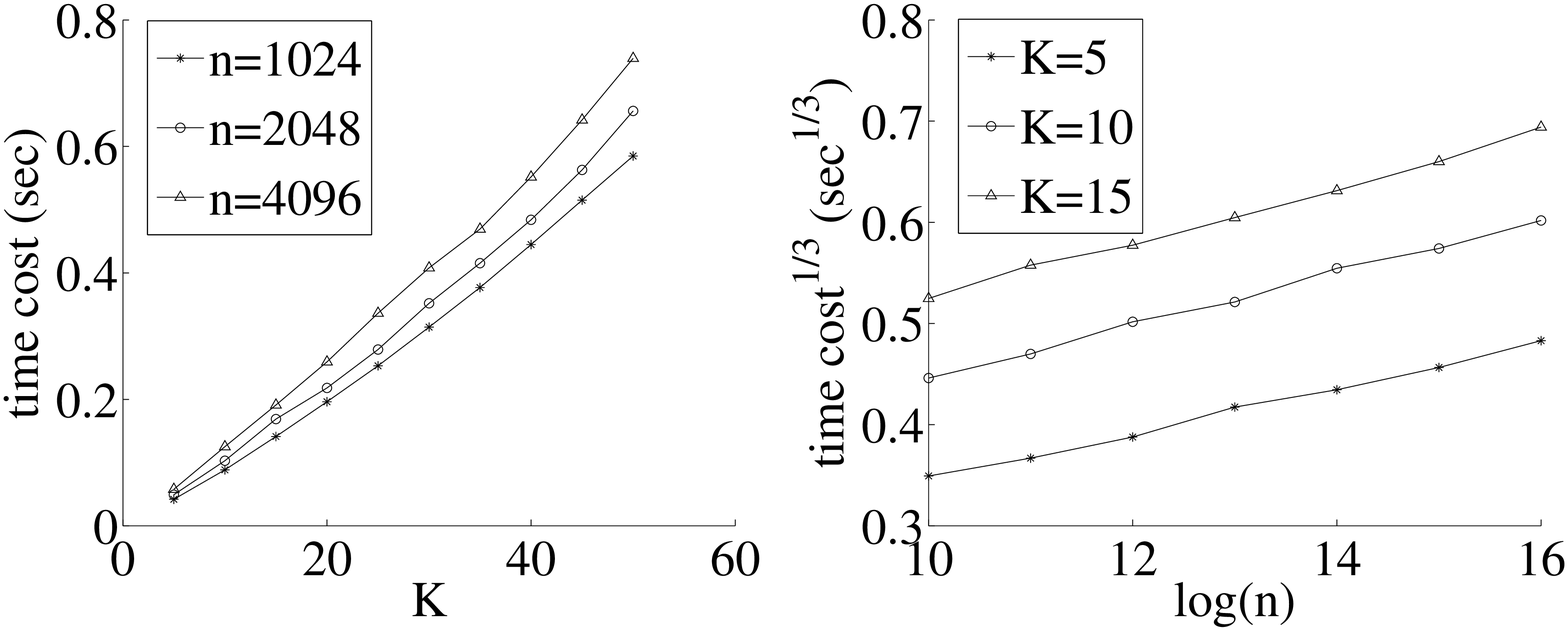}
\caption{Time cost. We choose $Q=2\log^2(n)$ and $\text{SNR}=20\text{dB}$. Different values of $n$ and $K$ are tested, and for each set of parameters, 100 experiments are conducted and the average time cost is shown.}
\label{fig_sim_time}
\end{figure}

In Figure \ref{fig_sim_snr}, we show the results of simulations on the probability of successful recovery as a function of the number of measurements and the SNR. Since the total number of measurements is dominated by $Q$. The sample complexity is mainly determined by $Q$. Therefore, we fixed $P$, i.e., the size of the test matrix, and tried different values of $Q$. 
From the results, we can see that the sublinear scheme can successfully recover the signal at relatively low SNR, such as 16dB, when $Q=2\log^2(n)$ and the number of measurements is $6.8\times 10^5$.

In Figure \ref{fig_sim_time}, we show the results of simulations on the time cost of the sublinear scheme\footnote{The simulations are conducted on a laptop with 2.8 GHz Intel Core i7 CPU and 16 GB memory.}. It can be seen that the time cost of sublinear scheme is indeed low and linear in $K$ and $\Theta(\log^3(n))$.

\section*{Acknowledgment}

The authors would like to thank Yudong Chen and Xiao Li for helpful discussions. 

\bibliographystyle{IEEEtran}
\bibliography{IEEEabrv,refs,ref_new}
\newpage
\appendix
\section*{Appendix}
\section{Notations}
We introduce some useful notations for the proofs. Here, $\FBNL{\cdot}$ denotes the Frobenius norm of a matrix, 
$\OPNL{\cdot}$ denotes the operator norm of a matrix.
For a sub-exponential random variable, $\SEPL{\cdot}$ denotes the sub-exponential norm of it; 
for a sub-gaussian random variable, $\SGSL{\cdot}$ denotes the sub-gaussian norm of it \cite{nonasym}.
The notations $c$, $c_i$, $C$, and $C_i$ represent absolute constants with positive value. 

In our model, we also assume that the noise $w_i$ satisfies $\EXPL{\ABS{w_i}}=\mu$, $\EXPL{w_i^2}=\sigma^2$, and $\SEPL{w_i} = \nu$.
Since the entries in $\mat{A}_0$ and $\mat{F}_0$ are bounded and thus sub-gaussian, we let $\eta=\SGSL{\ABS{a_{ij}}}$ 
and $\eta_0=\SGSL{\ABS{f_{0,ij}}}$, where $a_{ij}$ and $f_{0,ij}$ are entries of $\mat{A}_0$ and $\mat{F}_0$.

\section{Proof of Theorem \ref{thm_exh_overall}}\label{prf_exh_overall}

In order to prove Theorem \ref{thm_exh_overall}, we need to prove Lemma \ref{lem:energy_test} first. Here we restate Lemma \ref{lem:energy_test} with more details.
\renewcommand\thelemma{1}
\begin{lemma}\label{the_lemma_1}
There exists $\zeta>0$, determined by $\eta$, $\nu$, and $\sigma$, such that when $\phi>\mu/\zeta$, for any $t_0\in(\mu, \zeta\phi)$,
\begin{equation}\label{exh_test1}
\PRO{ \frac{1}{P}\ONEN{\vect{w}}\ge t_0}
= \BIGO(1/n^2),
\end{equation}
and
\begin{equation}\label{exh_test2}
\PRO{ \frac{1}{P}\ONEN{\vect{y}-\LNR(\hat{\vect{x}}\hat{\vect{x}}\HET)}< t_0}
= \BIGO(1/n^2),
\end{equation}
when $\hat{\vect{x}}\nsim\vect{x}$.
\end{lemma}

See the proof of Lemma \ref{lem:energy_test} in Appendix \ref{prf_lem_energy}. Now we can analyze the failure probability of the almost-linear scheme. Since the bipartite graph is $d$-left regular, there are $dn$ balls in all the bins. In the first iteration, we need to search all the balls. For each ball, we need to check $\Theta(1)$ possible magnitudes. Therefore, we need to do $\Theta(n)$ tests in the first iteration ($d$ is also a constant). Similarly, in the later iterations, we need to do at most $\Theta(n)$ tests. Since it is proved in \cite{Phasecode} that the number of iterations is a constant, we need to do $N_t=\Theta(n)$ tests.
Lemma \ref{lem:energy_test} tells us that, for any energy test, if the tests before it are all correct, and thus we have the correct colored balls, then, the error probability of this test is
$\BIGO(1/n^2)$. More specifically, let $E_i$ be the event that there is an error in the $i$th test, while the tests $1,\ldots,i-1$ are all correct. The event $E_{\text{test}}$ that there is error in all the energy tests can be decomposed as
$$
E_{\text{test}}=\bigcup_{i=1}^{N_t}{E_i}.
$$
By union bound, we have
$$
\PRO{E_{\text{test}}}\le N_t\sum_{i=1}^{N_t}\PRO{E_i}=\Theta(n)\BIGO(1/n^2)=\BIGO(1/n).
$$
Another possibility of making an error lies in the ball coloring algorithm itself. When there is no error in energy tests, this probability is $\BIGO(1/K)$ as analyzed in the noiseless case. Therefore the failure probability of the almost-linear scheme is
\begin{align}
\PRO{E_a}&=\PRO{E_a|E_{\text{test}}} \PRO{E_{\text{test}}} + \PROL{E_a|E_{\text{test}}^\complement} \PROL{E_{\text{test}}^\complement} \nonumber\\
&\le \PRO{E_{\text{test}}}+\PROL{E_a|E_{\text{test}}^\complement} \nonumber\\
&= \PRO{E_{\text{test}}}+\PRO{E_{\text{coloring}}} \nonumber\\
&=\BIGO(1/n)+\BIGO(1/K) \nonumber\\
&=\BIGO(1/K) \nonumber
\end{align}
The sample and computational complexity are already analyzed in Section \ref{sec:exh_search}. Now we complete the proof.

\section{Proof of Lemma \ref{the_lemma_1}}\label{prf_lem_energy}

To prove equation (\ref{exh_test1}), we simply use the Bernstein's inequality in \cite{nonasym}. For any $t>0$,
\begin{align}
&\PRO{\frac{1}{P}\sum_{i=1}^P(\ABS{w_i}-\EXP{\ABS{w_i}})>t}	 \nonumber\\
\le& \exp\left[-C_1P\min\left\{ \frac{t^2}{\nu^2},\frac{t}{\nu} \right\} \right]. \nonumber
\end{align}
Therefore, by choosing $t_0>\EXP{\ABS{w_i}}=\mu$ and $t=t_0-\mu$, we have
$$
\PRO{\frac{1}{P}\ONEN{\vect{w}}\ge t_0}\le \exp\left[ -\delta_1 P \right].
$$
Since $\delta_1$ is a constant and $P=\Theta(\log(n))$, equation (\ref{exh_test1}) is proved.

Now we prove equation (\ref{exh_test2}). Before getting into the details of the proof, we give the definition of a new notation $\phi$. For two vectors $\vect{p}, \vect{q}\in\SET^n$, it is easy to see that $\vect{p}\nsim\vect{q}\Leftrightarrow\vect{p}\vect{p}\HET-\vect{q}\vect{q}\HET\neq0$.
Since the entries of $\vect{p}$ and $\vect{q}$ lie in the quantized set $\SET$,
we know that there exists $\phi>0$, such that $\FBNL{\vect{p}\vect{p}\HET-\vect{q}\vect{q}\HET}>\phi$,
when $\vect{p}\nsim\vect{q}$, where $\phi$ depends on $\varepsilon$, $L_m$, and $L_p$. 
Then, we need the following lemma.

\renewcommand\thelemma{3}
\begin{lemma}\label{RIP}
Given two vectors $\vect{x}_1,\vect{x}_2\in\CMP^N$, let $\mat{X}=\vect{x}_1\vect{x}_1\HET-\vect{x}_2\vect{x}_2\HET\neq0$. $\LNR$ is the linear function defined in (\ref{lineardef}), and $\vect{w}$ is the noise. Then, for any $s>0$, we have,
\begin{align}
&\PRO{ \frac{1}{P}\ONEN{\LNR(\mat{X})+\vect{w}}< (\zeta-s\eta_d)\FBN{\mat{X}}-2s\nu } \nonumber\\
\le &\exp\left[ -C_0P\min{\{s^2,s\}} \right], \nonumber
\end{align}
where $\zeta>0$ depends on $\eta$, $\sigma$, and $\nu$, $\eta_d>0$ only depends on $\eta$.
\end{lemma}
See the proof of Lemma $\ref{RIP}$ in Appendix $\ref{prf_RIP}$. Note that $\vect{y}-\LNR(\hat{\vect{x}}\hat{\vect{x}}\HET)=\LNR(\vect{x}\vect{x}\HET-\hat{\vect{x}}\hat{\vect{x}}\HET)+\vect{w}$, and that $\FBNL{\vect{x}\vect{x}\HET-\hat{\vect{x}}\hat{\vect{x}}\HET}>\phi$. Now using Lemma $\ref{RIP}$, conditioning on $\vect{h}$, we have for any $s>0$,
\begin{align}
&\PRO{ \frac{1}{P}\ONEN{\vect{y}-\LNR(\hat{\vect{x}}\hat{\vect{x}}\HET)}< \zeta\phi-(\eta_d\phi+2\nu)s \ |\ \vect{h} } \nonumber\\
\le &\exp\left[ -C_0P\min{\{s^2,s\}} \right]. \label{rip_s}
\end{align}
Since (\ref{rip_s}) holds for any $\vect{h}$, we know that it also holds without conditioning on $\vect{h}$. If $\zeta\phi>t_0$, we can choose $s=\frac{\zeta\phi-t_0}{\eta_d\phi+2\nu}$, then
$$
\PRO{ \frac{1}{P}\ONEN{\vect{y}-\LNR(\hat{\vect{x}}\hat{\vect{x}}\HET)}< t_0}\le \exp\left[ -\delta_2P \right].
$$
Since $\delta_2$ is a constant and $P=\Theta(\log(n))$, equation (\ref{exh_test2}) is proved.

Then, we can conclude that there exists $\zeta$, determined by the statistics of noise, such that when $\phi>\mu/\zeta$, for any threshold $t_0\in(\mu, \zeta\phi)$,
the energy test fails with probability $\BIGO(1/n^2)$. This completes the proof of Lemma \ref{the_lemma_1}. 

\section{Proof of Lemma \ref{RIP}}\label{prf_RIP}
The proof of Lemma \ref{RIP} is based on similar ideas appeared in \cite{chen2014convex}. 
Let $\vect{\xi}=\LNR(\mat{X})+\vect{w}$, then $\xi_i=\vect{a}_i\HET \mat{X} \vect{a}_i+w_i$. 
According to the definition of the matrix $\mat{A}$, we know that the Hanson-Wright inequality for complex random variables in Appendix \ref{prf_comp_hw} holds for $\vect{a}_i\HET \mat{X} \vect{a}_i$ and we have for every $t>0$,
\begin{align}
&\PRO{\ABS{\vect{a}_i\HET \mat{X} \vect{a}_i-\EXP{\vect{a}_i\HET \mat{X} \vect{a}_i}}>t} \nonumber \\
\le & 6\exp{\left[-c\min\left\{ \frac{t^2}{\eta^4\FBN{\mat{X}}^2}, \frac{t}{\eta^2\OPN{\mat{X}}} \right\}\right]} \nonumber \\
\le & 6\exp{\left[-c\min\left\{ \frac{t}{\eta^2\FBN{\mat{X}}}-\frac{1}{4}, \frac{t}{\eta^2\FBN{\mat{X}} }\right\}\right]} \nonumber\\
\le & 6\exp{\left[c\left(\frac{1}{4}-\frac{t}{\eta^2\FBN{\mat{X}}}\right)\right]}, \nonumber
\end{align}
where the second inequality is due to the fact that $(a-1/2)^2\ge 0$ and $\|\mat{X}\|\le \FBNL{\mat{X}}$. From \cite{nonasym}, we know that $\vect{a}_i\HET \mat{X} \vect{a}_i-\EXP{\vect{a}_i\HET \mat{X} \vect{a}_i}$ is a sub-exponential random variable with sub-exponential norm 
\begin{equation}\label{subexp1}
\SEP{\vect{a}_i\HET \mat{X} \vect{a}_i-\EXP{\vect{a}_i\HET \mat{X} \vect{a}_i}}\le C\eta^2\FBN{\mat{X}}.
\end{equation}
On the other hand, 
$$
\ABS{\EXP{\vect{a}_i\HET \mat{X} \vect{a}_i}}=\frac{1}{2}\ABS{\TWON{\vect{x}_1}^2-\TWON{\vect{x}_2}^2}\le\frac{1}{2}\FBN{\mat{X}}.
$$
Thus, it gives us
\begin{align}
\SEP{\xi_i} =& \SEP{\vect{a}_i\HET \mat{X} \vect{a}_i-\EXP{\vect{a}_i\HET \mat{X} \vect{a}_i} + \EXP{\vect{a}_i\HET \mat{X} \vect{a}_i} + w_i}  \nonumber\\
\le&\SEP{\vect{a}_i\HET \mat{X} \vect{a}_i-\EXP{\vect{a}_i\HET \mat{X} \vect{a}_i}} + \ABS{\EXP{\vect{a}_i\HET \mat{X} \vect{a}_i}} +\nu \nonumber\\
 \le& (C\eta^2+1/2)\FBN{\mat{X}}+\nu, \label{subexp2}
\end{align}
where the first inequality is due to the fact that $\EXP{\vect{a}_i\HET \mat{X} \vect{a}_i}$ is a constant, $\SEPL{\EXP{\vect{a}_i\HET \mat{X} \vect{a}_i}} = \ABSL{\EXP{\vect{a}_i\HET \mat{X} \vect{a}_i}}$, and that $\SEPL{w_i} = \nu$. Then
\begin{equation}\label{subexp3}
\SEP{\ABS{\xi_i}-\EXP{\ABS{\xi_i}}} \le 2\SEP{\xi_i}\le \eta_d\FBN{\mat{X}}+2\nu,
\end{equation}
where $\eta_d=2C\eta^2+1$. Now according to Bernstein's inequality in \cite{nonasym}, we have for every $t>0$,
\begin{align}
&\PRO{ \frac{1}{P}\sum_{i=1}^P(\ABS{\xi_i}-\EXP{\ABS{\xi_i}})<-t }  \nonumber\\
\le &\exp\left[-C_0P\min\left\{\frac{t^2}{(\eta_d\FBN{\mat{X}}+2\nu)^2},\frac{t}{\eta_d\FBN{\mat{X}}+2\nu}\right\}\right]. \nonumber
\end{align}
Let $t=s(\eta_d\FBN{\mat{X}}+2\nu)$. For any $s>0$,
\begin{align}
&\PRO{ \frac{1}{P}\sum_{i=1}^P(\ABS{\xi_i}-\EXP{\ABS{\xi_i}})<-s(\eta_d\FBN{\mat{X}}+2\nu) } \nonumber\\
\le &\exp\left[ -C_0P\min{\{s^2,s\}} \right].	\label{Bernstein}
\end{align}
By Cauchy-Schwartz inequality,  for any $i\in[P]$, we have 
$$
\left( \EXP{\xi_i^2} \right)^2\le \EXP{\ABS{\xi_i}}\EXP{\ABS{\xi_i}^3} \le \EXP{\ABS{\xi_i}}\sqrt{\EXP{\xi_i^2}\EXP{\xi_i^4}},
$$
which implies
\begin{equation}\label{expbound0}
\EXP{\ABS{\xi_i}}\ge\sqrt{\frac{(\EXP{\xi_i^2})^3}{\EXP{\xi_i^4}}}.
\end{equation}
According to the definition of sub-exponential norm and the fact that $\eta_d>1$, we have
\begin{align}
\EXP{\xi_i^4}&\le(4\SEP{\xi_i})^4\le(2\eta_d\FBN{\mat{X}}+4\nu)^4  \nonumber\\
&\le (8\eta_d^2\FBN{\mat{X}}^2+32\nu^2)^2. \label{expbound1}
\end{align}
On the other hand, we have
\begin{align}
\EXP{\xi_i^2}&=\EXP{(\vect{a}_i\HET \mat{X} \vect{a}_i)^2}+\EXP{w_i^2}  \nonumber\\
&=\EXP{(\vect{a}_i\HET \mat{X} \vect{a}_i)\TR{\vect{a}_i\vect{a}_i\HET\mat{X}}}+\sigma^2 \nonumber\\
&=\EXP{\TR{(\vect{a}_i\HET \mat{X} \vect{a}_i)\vect{a}_i\vect{a}_i\HET\mat{X}}}+\sigma^2 \nonumber\\ 
&=\TR{\EXP{(\vect{a}_i\HET \mat{X} \vect{a}_i)\vect{a}_i\vect{a}_i\HET}\mat{X}}+\sigma^2 \nonumber\\ 
&=\frac{1}{4}\TR{(\mat{X}+\TR{\mat{X}}\mat{I})\mat{X}}+\sigma^2 \label{inter1} \\ 
&\ge \frac{1}{4}\FBN{\mat{X}}^2+\sigma^2. \label{expbound2}
\end{align}
Here we give an explanation of equation (\ref{inter1}). Let $\mat{Y}=(\vect{a}_i\HET \mat{X} \vect{a}_i)\vect{a}_i\vect{a}_i\HET$. Then,
\begin{align}
&\EXP{Y_{jk}} \nonumber \\
=& \EXP{\sum_{1\le g,h\le n}a_{ig}^*X_{gh}a_{ih}a_{ij}a_{ik}^*}  \nonumber\\
=& \sum_{g=1}^n\EXP{a_{ig}^*X_{gg}a_{ig}a_{ij}a_{ik}^*} + \sum_{g\neq h}\EXP{a_{ig}^*X_{gh}a_{ih}a_{ij}a_{ik}^*}.   \nonumber
\end{align}
If $j=k$, we have
\begin{align}
\EXPL{Y_{jj}}&=\sum_{g=1}^n X_{gg}\EXPL{\ABSL{a_{ig}}^2\ABSL{a_{ij}}^2} + \sum_{g\neq h} X_{gh}\EXPL{a_{ig}^*a_{ih}\ABSL{a_{ij}}^2}  \nonumber\\
&=\frac{1}{4}(\TR{X}+X_{jj}). \nonumber
\end{align}
If $j \neq k$, we have
\begin{align}
\EXPL{Y_{jk}}&=\sum_{g=1}^n X_{gg}\EXPL{\ABSL{a_{ig}}^2a_{ij}a_{ik}^*} + \sum_{g\neq h} X_{gh}\EXPL{a_{ig}^*a_{ih}a_{ij}a_{ik}^*}  \nonumber\\
&= X_{jk} \EXPL{\ABSL{a_{ij}}^2\ABSL{a_{ik}}^2}  \nonumber\\
&=\frac{1}{4}X_{jk}. \nonumber
\end{align}
Therefore, $\EXPL{\mat{Y}}=\frac{1}{4}(\mat{\mat{X}}+\TR{\mat{X}}\mat{I})$.

By combining (\ref{expbound0}), (\ref{expbound1}), and (\ref{expbound2}), we have
\begin{align}
\EXP{\ABS{\xi_i}}&\ge\sqrt{\left(\frac{\frac{1}{4}\FBN{\mat{X}}^2+\sigma^2}{8\eta_d^2\FBN{\mat{X}}^2+32\nu^2}\right)^2\left(\frac{1}{4}\FBN{\mat{X}}^2+\sigma^2\right)} \nonumber\\
&\ge \zeta\FBN{\mat{X}}, \nonumber
\end{align}
where $\zeta=\frac{1}{2}\min\left\{\frac{1}{32\eta_d^2},\frac{\sigma^2}{32\nu^2}\right\}$ is a constant determined by the distribution of $a_{ij}$ and $w_i$.
Then according to (\ref{Bernstein}), we have
\begin{align}
&\PRO{ \frac{1}{P}\sum_{i=1}^P\ABS{\xi_i}< \zeta\FBN{\mat{X}}-s(\eta_d\FBN{\mat{X}}+2\nu) }  \nonumber\\
\le& \exp\left[ -C_0P\min{\{s^2,s\}} \right], \nonumber
\end{align}
which completes the proof.
\section{Proof of Theorem \ref{thm_fast_overall}}\label{prf_thm_fast}

To prove Theorem \ref{thm_fast_overall}, we need to use Lemma \ref{lem:fast_one_section}. Here, we restate Lemma \ref{lem:fast_one_section}, providing more details.
\renewcommand\thelemma{2}
\begin{lemma}\label{the_lem_2}
If $T_s=1$, $\SUPP{\vect{x}_s}=\{l_s\}$, and threshold $t_1\in(0,\varepsilon^2/2)$, then for any $j\in[R]$,
$$
\PRO{\tilde{b}_j\neq b_{jl_s}}=\BIGO(1/K^3).
$$
\end{lemma}
See the proof of Lemma \ref{the_lem_2} in Appendix \ref{prf_lem_2}. Then, by union bound, we know that $\mathbb{P}\{\tilde{\vect{b}} \neq \vect{B}_{l_s}\}=\BIGO(R/K^3)\le\BIGO(1/K^2)$, since $K=\beta n^\delta$. Now we can see that we can reliably find $l_s$ from the measurements with probability $1-\BIGO(1/K^2)$. For a bin with $T_s=1$, the probabilities of error in index tests and energy test are $\BIGO(1/K^2)$ and $\BIGO(1/n^2)$, respectively. Therefore, the error probability of the tests for this bin is $\BIGO(1/K^2)$. For a bin with $T_s>1$, only the energy test should be considered and its error probability is $\BIGO(1/n^2)$. Then, we know the probability of making mistakes in the index and energy tests is $\BIGO(1/K^2)$. Since there are $\Theta(K)$ bins and a constant number of iterations, using the same decomposition method as in the proof of Theorem \ref{thm_exh_overall}, we know that the error probability of all the tests is $\BIGO(1/K)$. Similar to the almost-linear scheme, considering the $\BIGO(1/K)$ probability of incomplete recovery in the ball coloring algorithm when there is no error in the index and energy tests, the failure probability of sublinear scheme is $\PROL{E_s}=\BIGO(1/K)$. Since the sample and computational complexity are already analyzed in Section \ref{sec:fast_search}, the proof of Theorem \ref{thm_fast_overall} is now complete.

\section{Proof of Lemma \ref{lem:fast_one_section}}\label{prf_lem_2}
First, we define an event $E_h$ such that there are more than $C_3\log K$ active balls in a bin. As mentioned in \cite{pawar2013pulse}, we have $\PROL{E_h}=\BIGO(1/K^3)$. Now we condition on the coding pattern $\vect{h}$ such that $E_h^\complement$ happens, and thus $\ABS{\SUPP{\vect{x}}}=T\le C_3\log K$.
Similar to the almost-linear algorithm, we define $R+1$ linear mappings, $\LNR_0, \LNR_1, \ldots, \LNR_R$, where
$$
\LNR_0:\ \mat{X}\mapsto \{\vect{a}_i\HET \mat{X} \vect{a}_i\}_{i\in[P]},
$$
$$
\LNR_j:\ \mat{X}\mapsto \{\vect{f}_{j,i}\HET \mat{X} \vect{f}_{j,i}\}_{i\in[Q]}, \ {\rm for\ } j\in[R].
$$
Then, We have $\vect{y}_j=\LNR_j(\vect{x}\vect{x}\HET)+\vect{w}_j$, $j\in\{0\}\cup[R]$.

Define the matrix $\tilde{\mat{X}}=\{\tilde{X}_{ij}\}_{N\times N}:=\vect{x}\vect{x}\HET-\tilde{\vect{x}}_c\tilde{\vect{x}}_c\HET=\vect{x}\vect{x}\HET-\vect{x}_c\vect{x}_c\HET$. 
Then $\tilde{\vect{y}}_j=\LNR_j(\tilde{\mat{X}})+\vect{w}_j$.
There is $\tilde{y}_{j,i}=\vect{f}_{j,i}\HET \tilde{\mat{X}} \vect{f}_{j,i}+w_{j,i}$.
Let $f_{j,i,m}$ be the $m$th element of $\vect{f}_{j,i}$. Since for a fixed $j$, $f_{j,i,m}$'s are independent, using the similar argument in Appendix \ref{prf_RIP}, we know that 
$$
\SEP{\vect{f}_{j,i}\HET \tilde{\mat{X}} \vect{f}_{j,i} - \EXP{\vect{f}_{j,i}\HET \tilde{\mat{X}} \vect{f}_{j,i}}} \le C_2\eta_0^2\FBN{\tilde{\mat{X}}}.
$$
Then we have $\SEPL{\tilde{y}_{j,i}-\EXPL{\tilde{y}_{j,i}}}\le C_2\eta_0^2\FBNL{\tilde{\mat{X}}} + \nu$.
Since there are $2T-1$ nonzero entries in $\tilde{\mat{X}}$, we have $\FBNL{\tilde{\mat{X}}}\le \sqrt{2T-1}L_m\varepsilon$. 
We also have $T\le C_3\log K$, therefore we have
$\SEPL{\tilde{y}_{j,i}-\EXPL{\tilde{y}_{j,i}}}\le C_4\eta_0^2L_m\varepsilon \sqrt{\log K}+\nu\le \zeta_0\sqrt{\log K}$, where $\zeta_0$ is determined by $\eta_0$, $L_m$, $\varepsilon$, and $\nu$.

On the other hand, since $T_s=1$ and $\SUPP{\vect{x}_s}=\{l_s\}$, $\tilde{\mat{X}}$ only has one nonzero element on the diagonal, i.e.,
$\tilde{X}_{l_sl_s}=\ABSL{x_{l_s}}^2$. We have $\EXPL{\tilde{y}_{j,i}}=\EXPL{\ABSL{f_{j,i,l_s}}^2}\ABSL{x_{l_s}}^2=b_{jl_s}\ABSL{x_{l_s}}^2$. According to Bernstein's inequality, for every $t\ge0$, we have
\begin{align}
&\PRO{\ABS{\frac{1}{Q}\sum_{i=1}^Q (\tilde{y}_{j,i} - b_{jl_s}\ABS{x_{l_s}}^2 )}>t\ |\ \vect{h}}  \nonumber\\
\le &2\exp \left[ -C_5 Q \min \left\{ \frac{t^2}{\zeta_0^2\log K}, \frac{t}{\zeta_0\sqrt{\log K}} \right\} \right] \nonumber\\
\le &2\exp \left[ -\frac{C_5}{\zeta_0^2} \sqrt{Q} \min\left\{ t^2,t \right\} \right], \nonumber
\end{align}
where the last inequality comes from the fact that $Q=\Theta(\log^2N)$. Now choose $t_1=t<\varepsilon^2/2$. When $b_{jl_s}=0$, we know that 
\begin{align}
&\PRO{\ABS{\frac{1}{Q}\sum_{i=1}^Q \tilde{y}_{j,i}}>t_1\ |\ \vect{h}}  \nonumber\\
\le &2\exp \left[ -\frac{C_5}{\zeta_0^2} \sqrt{Q} \min\left\{ t_1^2,t_1 \right\} \right],  \label{fast_test1}
\end{align}
and when $b_{jl_s}=1$, we have
\begin{align}
&\PRO{\ABS{\frac{1}{Q}\sum_{i=1}^Q \tilde{y}_{j,i}}<t_1\ |\ \vect{h}}    \nonumber\\
\le &\PRO{\frac{1}{Q}\sum_{i=1}^Q \tilde{y}_{j,i}<t_1\ |\ \vect{h}}  \nonumber\\
\le &\PRO{\frac{1}{Q}\sum_{i=1}^Q \tilde{y}_{j,i}<\ABS{x_{l_s}}^2-t_1\ |\ \vect{h}}  \label{interm2} \\
\le &\PRO{\ABS{\frac{1}{Q}\sum_{i=1}^Q \tilde{y}_{j,i}-\ABS{x_{l_s}}^2}>t_1\ |\ \vect{h}}    \nonumber\\
\le &2\exp \left[ -\frac{C_5}{\zeta_0^2} \sqrt{Q} \min\left\{ t_1^2,t_1 \right\} \right],   \label{fast_test2}
\end{align}
where the inequality (\ref{interm2}) is due to the fact that $t_1<\varepsilon^2/2$ and $\ABS{x_{l_s}}^2\ge \varepsilon^2$. Define the error events $E_{\text{index}}=\{\ABSL{\frac{1}{Q}\sum_{i=1}^Q \tilde{y}_{j,i}}>t_1\}$, when $b_{jl_s}=0$, and $E_{\text{index}}=\{\ABSL{\frac{1}{Q}\sum_{i=1}^Q \tilde{y}_{j,i}}<t_1\}$, when $b_{jl_s}=1$.
Then, since $Q=\Theta(\log^2(n))$ and inequalities (\ref{fast_test1}) and (\ref{fast_test2}) hold for any $\vect{h}\in E_h^\complement$, we have,
$$
\PROL{E_{\text{index}}|E_h^\complement}=\BIGO(1/K^3).
$$
Now we know that
\begin{align}
\PROL{E_{\text{index}}}&=\PROL{E_{\text{index}}|E_h^\complement}\PROL{E_h^\complement}+\PROL{E_{\text{index}}|E_h}\PROL{E_h} \nonumber\\
&\le \PROL{E_{\text{index}}|E_h^\complement} + \PROL{E_h} \nonumber\\
& = \BIGO(1/K^3)+\BIGO(1/K^3) \nonumber\\
& = \BIGO(1/K^3), \nonumber
\end{align}
which completes the proof.

\section{Hanson-Wright Inequality for Complex Random Variables}\label{prf_comp_hw}
\begin{theorem}\label{thm_comp_hw}
Let $\vect{\gamma}=\{\gamma_i\}_{i\in[n]}\in\CMP^n$ be a random vector with independent entries $\gamma_i$, satisfying 
$\EXP{\gamma_i}=0$, and $\ABS{\gamma_i}$ is sub-gaussian with $\SGS{\ABS{\gamma_i}}\le\eta$ for all $i\in[n]$. 
Let $\mat{U}\in\CMP^{n\times n}$ be a Hermitian matrix. Then, for every $t\ge 0$,
\begin{align}
&\PRO{\ABS{\vect{\gamma}\HET \mat{U} \vect{\gamma}-\EXP{\vect{\gamma}\HET \mat{U} \vect{\gamma}}}>t} \nonumber\\
\le&6\exp{\left[-c_0\min\left\{\frac{t^2}{\eta^4\FBN{\mat{U}}^2}, \frac{t}{\eta^2\OPN{\mat{U}}}\right\}\right]}. \nonumber
\end{align}
\end{theorem}
\begin{proof}
Let $\vect{\alpha}=\{\alpha_i\}_{i\in[n]}$ and $\vect{\beta}=\{\beta_i\}_{i\in[n]}$ be the real and imaginary parts of $\vect{\gamma}$.
Then, we know that $\alpha_i$'s and $\beta_i$'s are sub-gaussian random variables with $\SGS{\alpha_i}\le \eta$ and 
$\SGS{\beta_i}\le \eta$ for all $i\in[n]$.
Note that here, although $\gamma_i$'s are independent, the real and imaginary parts of $\gamma_i$ are not necessarily independent for a certain $i$. In other words, for any $i$, $\alpha_i$ and $\beta_i$ may not be independent.

Let $\mat{V}$ and $\mat{W}$ be the real and imaginary parts of $\mat{U}$. 
Since $\mat{U}$ is a Hermitian matrix, we have $\mat{V}=\mat{V}\TSP$ and $\mat{W}=-\mat{W}\TSP$. We also know that 
$\vect{\gamma}\HET \mat{U} \vect{\gamma}$ is a real number. Then, we have
$$
\vect{\gamma}\HET \mat{U} \vect{\gamma} = \vect{\alpha}\TSP \mat{V} \vect{\alpha} - 2\vect{\alpha}\TSP\mat{W}\vect{\beta}+\vect{\beta}\TSP\mat{V}\vect{\beta}.
$$
Therefore, $\PRO{\ABS{\vect{\gamma}\HET \mat{U} \vect{\gamma}-\EXP{\vect{\gamma}\HET \mat{U} \vect{\gamma}}}>t}$ is upper bounded by three terms,
\begin{align}
&\PRO{\ABS{\vect{\gamma}\HET \mat{U} \vect{\gamma}-\EXP{\vect{\gamma}\HET \mat{U} \vect{\gamma}}}>t} \nonumber\\ 
\le &\PRO{\ABS{\vect{\alpha}\TSP \mat{V} \vect{\alpha}-\EXP{\vect{\alpha}\TSP \mat{V} \vect{\alpha}}}>t/4}  \nonumber\\ 
+ &\PRO{\ABS{\vect{\alpha}\TSP\mat{W}\vect{\beta}-\EXP{\vect{\alpha}\TSP\mat{W}\vect{\beta}}}>t/4} \nonumber\\
+ &\PRO{\ABS{\vect{\beta}\TSP\mat{V}\vect{\beta}-\EXP{\vect{\beta}\TSP\mat{V}\vect{\beta}}}>t/4}. \label{separating}
\end{align}
Since $\alpha_i$'s are independent and $\EXP{\alpha_i}=0$, according to the Hanson-Wright inequality for real numbers\cite{HansonWright}, we have
\begin{align}
&\PRO{\ABS{\vect{\alpha}\TSP \mat{V} \vect{\alpha}-\EXP{\vect{\alpha}\TSP \mat{V} \vect{\alpha}}}>t/4}  \nonumber\\
\le&2\exp{\left[-c_1\min\left\{\frac{t^2}{\eta^4\FBN{\mat{V}}^2}, \frac{t}{\eta^2\OPN{\mat{V}}}\right\}\right]}. \nonumber
\end{align}
We also have $\FBN{\mat{V}}\le\FBN{\mat{U}}$, $\OPN{\mat{V}}\le\OPN{\mat{U}}$. Therefore,
\begin{align}
&\PRO{\ABS{\vect{\alpha}\TSP \mat{V} \vect{\alpha}-\EXP{\vect{\alpha}\TSP \mat{V} \vect{\alpha}}}>t/4}    \nonumber\\
\le&2\exp{\left[-c_1\min\left\{\frac{t^2}{\eta^4\FBN{\mat{U}}^2}, \frac{t}{\eta^2\OPN{\mat{U}}}\right\}\right]}.  \label{alpha_part}
\end{align}
And similarly,
\begin{align}
&\PRO{\ABS{\vect{\beta}\TSP \mat{V} \vect{\beta}-\EXP{\vect{\beta}\TSP \mat{V} \vect{\beta}}}>t/4}	 \nonumber\\
\le&2\exp{\left[-c_2\min\left\{\frac{t^2}{\eta^4\FBN{\mat{U}}^2}, \frac{t}{\eta^2\OPN{\mat{U}}}\right\}\right]}. \label{beta_part}
\end{align}

Now consider the cross term. Let $W_{ij}$ be the entries of $\mat{W}$. Since $\mat{W}=-\mat{W}\TSP$, $W_{ii}=0$ for all $i\in[n]$. 
Then we have $\vect{\alpha}\TSP\mat{W}\vect{\beta}=\sum_{i\neq j}{W_{ij}\alpha_i\beta_j}$, 
and $\EXP{\vect{\alpha}\TSP\mat{W}\vect{\beta}}=0$. Then, we can bound $\PRO{\ABS{\vect{\alpha}\TSP\mat{W}\vect{\beta}}>t/4}$ in the same way as in \cite{HansonWright}. We have
\begin{align}
&\PRO{\ABS{\vect{\alpha}\TSP\mat{W}\vect{\beta}}>t/4} \nonumber\\
\le&2\exp{\left[-c_3\min\left\{ \frac{t^2}{\eta^4\FBN{\mat{U}}^2}, \frac{t}{\eta^2\OPN{\mat{U}}} \right\}\right]}.  \label{cross_answer}
\end{align}
By combining (\ref{alpha_part}), (\ref{beta_part}), and (\ref{cross_answer}), Theorem \ref{thm_comp_hw} is proved.

\end{proof}
\end{document}